\documentclass[a4paper,12pt]{article}

\usepackage[english]{babel}
\usepackage[T1]{fontenc}

\usepackage[a4paper,top=3cm,bottom=2cm,left=3cm,right=3cm,marginparwidth=1.75cm]{geometry}
   
\usepackage{amsmath}
\usepackage{graphicx}
\usepackage{color}
\usepackage[colorinlistoftodos]{todonotes}
\usepackage[colorlinks=true, allcolors=blue]{hyperref}
\usepackage{subcaption}

\title{Massive MIMO in Sub-6 GHz and mmWave: Physical, Practical, and Use-Case Differences}

\date{}

\author{Emil Bj\"ornson\thanks{E. Bj\"ornson is with Link\"oping University, Link\"oping, Sweden. E-mail: emil.bjornson@liu.se.}, Liesbet Van der Perre\thanks{L. Van der Perre is with Katholieke Universiteit Leuven, Belgium and with Lund University, Sweden. E-mail: liesbet.vanderperre@kuleuven.be.}, Stefano Buzzi\thanks{S. Buzzi is with University of Cassino and Lazio Meridionale and with the Consorzio Nazionale Interuniversitario per le Telecomunicazioni (CNIT), Italy. E-mail: s.buzzi@unicas.it.}, and Erik G. Larsson\thanks{E. G. Larsson is with Link\"oping University, Link\"oping, Sweden. E-mail: erik.g.larsson@liu.se.}}

\begin{document}
\maketitle

\begin{abstract}
The use of base stations (BSs) and access points (APs) with a large number of antennas, called Massive MIMO (multiple-input multiple-output), is a key technology for increasing the capacity of 5G networks and beyond. While originally conceived for conventional sub-6\,GHz frequencies, Massive MIMO (mMIMO) is ideal also for frequency bands in the range 30-300\,GHz, known as millimeter wave (mmWave). Despite conceptual similarities, the way in which mMIMO can be exploited in these bands is radically different, due to their specific propagation behaviors and hardware characteristics. 
This paper reviews these differences and their implications, while dispelling common misunderstandings.
Building on this foundation, we suggest appropriate signal processing schemes and use cases to efficiently exploit mMIMO in both frequency bands.
\end{abstract}

\section*{Introduction}

mMIMO uses arrays with many antennas at the BS to provide vast signal amplification by beamforming and high spatial resolution to multiplex many simultaneous users.
Although small-scale MIMO technology has been around for decades, the practical gains have been modest due to the small number of antennas which seldom give sufficient spatial resolution to support many spatially multiplexed streams. 
mMIMO has been demonstrated to achieve an order-of-magnitude higher spectral efficiency
in real life, with practical acquisition of channel state information (CSI) \cite{Marzetta}. 3GPP is steadily increasing the maximum number of antennas in LTE and since 64 antennas are supported in Release 15, mMIMO has become an integral component of 5G.

Another key approach to increase the capacity of future wireless networks is the operation in mmWave bands. There are many GHz of unused spectrum above 30\,GHz, which can be used as a complement to the current sub-6\,GHz bands. The path-loss and blockage phenomena are more severe in mmWave bands, but can be (partially) overcome by keeping the same physical size of the antenna array as on lower frequencies, which is achieved by mMIMO. There are, however, fundamental differences between how mMIMO technology can be designed, implemented, and exploited in sub-6\,GHz and mmWave bands. In this paper, we provide a comparative overview, highlighting three main differences:

\begin{enumerate}
\item
{\bf The propagation channels} build on the same physics, but basic phenomena such as diffraction, attenuation, and Fresnel zones are substantially different.

\item
{\bf The hardware implementation} architecture changes with the increasing carrier frequency. More antennas can be integrated into a given area, but the insertion losses, intrinsic power-overhead in radio-frequency (RF) generation, and amplification result in diminishing gains.

\item
{\bf The signal processing algorithms} depend on propagation and hardware. Channel estimation is resource-demanding at sub-6\,GHz, while beamforming is straightforward. Conversely, mmWave channel estimation and beamforming are theoretically simpler since there are fewer propagation paths, but become challenging if hybrid beamforming is used.

\end{enumerate} 

In the remainder of this article, we elaborate on these differences, including that they manifest how sub-6\,GHz and mmWave bands are to be exploited to target different use-cases in 5G and beyond.

\section*{Difference I: The propagation channel}

An understanding of the electromagnetic propagation is crucial when considering mMIMO systems and frequencies up to mmWave bands.
The channels behave fundamentally different from what we are used to in cellular networks, which exposes weaknesses in the channel modeling simplifications conventionally made.

\subsection*{Sub-6\,GHz: favorable propagation and spatial correlation}

\begin{figure}
        \centering
        \begin{subfigure}[b]{\columnwidth} \centering 
                \includegraphics[width=.7\columnwidth]{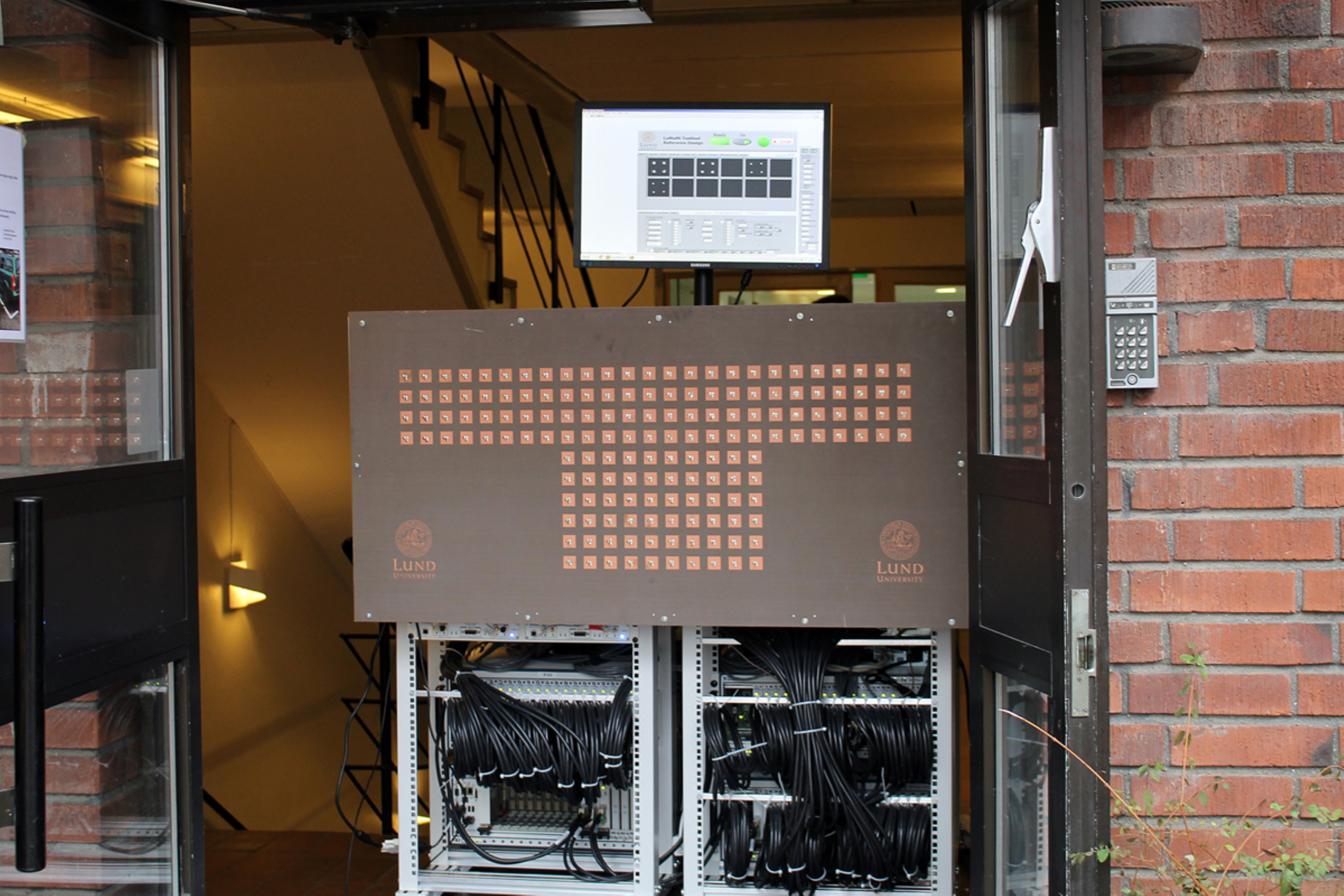} 
                \caption{Co-located mMIMO testbed at Lund University.}  
                \label{figure:testbed-Lund}
        \end{subfigure}  \vskip5mm
\begin{subfigure}[b]{\columnwidth} \centering 
                \includegraphics[width=.7\columnwidth]{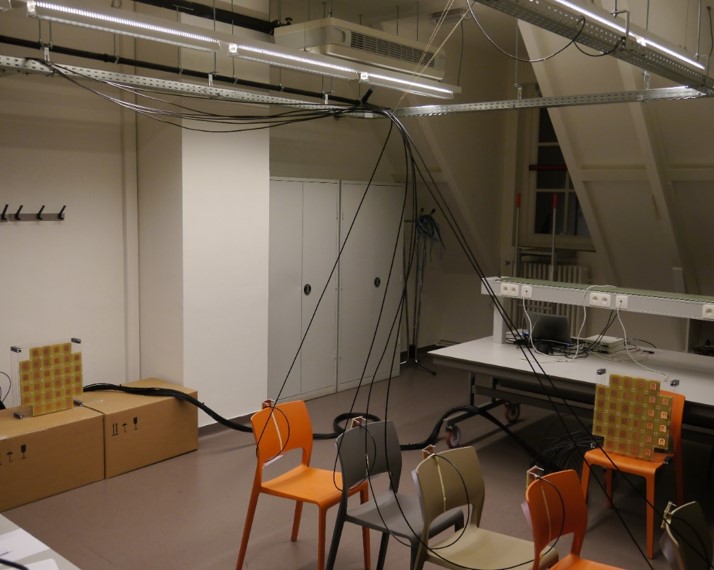} 
                \caption{Distributed mMIMO testbed at KU Leuven.} 
                \label{figure:testbed-KUL}
        \end{subfigure}
        
        \caption{Photos (courtesy U Lund - EIT and KU Leuven - ESAT) of two operational real-time mMIMO testbeds, both built using hardware components from National Instruments and using a 20\,MHz bandwidth. The co-located testbed at Lund University supports 100 BS antennas around a carrier frequency of 3.7\,GHz. The testbed at KU Leuven supports 64 BS antennas designed for the  2.4-2.62\,GHz and 3.4-3.6\,GHz bands and supports distributed operation.}
        \label{figure:testbed}  
\end{figure}

Radio channels below 6\,GHz have been widely studied for single-antenna and small-scale MIMO systems. The propagation depends on path-loss and shadowing, called large-scale fading, and multi-path propagation, resulting in small-scale fading.
In recent years, measurement campaigns have been carried out to characterize sub-6\,GHz mMIMO channels \cite{Gao2015,Harris2017a}. For example, the real-time testbed at Lund University, shown in Figure~\ref{figure:testbed-Lund}, has substantially contributed to the understanding of both mMIMO propagation phenomena and hardware implementation. Figure~\ref{figure:testbed-KUL} shows an alternative distributed mMIMO deployment. mMIMO measurements show that the UEs' channels become closer to orthogonal with an increasing number of antennas, referred to as \emph{favorable propagation} \cite{MLYN2016book}. Differently from small-scale MIMO, the large-scale fading can potentially vary significantly between the antennas in mMIMO. This occurs, for example, when a part of a physically large array is more shadowed than the rest \cite[Sec.~7.3]{massivemimobook} or when using cylindrical arrays where the antennas point in different directions.

Many researchers consider i.i.d.~Rayleigh fading channels in their assessment of mMIMO. This approach is analytically tractable,  provides insightful rate expressions, and leads to \emph{channel hardening}, where the impact of small-scale fading reduces as more antennas are added. In contrast, correlated fading channels are more complicated to analyze \cite{massivemimobook}.
However, in practice, not every channel is well-modeled as having i.i.d.~channel coefficients. Some UEs will have strong line-of-sight (LoS) components and many UEs will feature spatially correlated small-scale fading. These characteristics must be modeled to capture how spatial correlation leads to more (less) interference between UEs that have similar (different) spatial correlation characteristics \cite{massivemimobook}. The channel hardening effect is also weaker under spatial correlation.

Although the beamforming becomes more directive as the number of antennas $M$ increases, it has no effect on how frequently we need to re-estimate the channel under mobility. To show this, consider a UE in LoS that moves a fraction $\mu \leq 1/8$ of the wavelength. The $m$th channel coefficient is  phase-shifted by $e^{j 2\pi \phi_m}$, where $\phi_m \in [-\mu,\mu]$ depends on the direction of movement. If the beamforming is fixed at the original UE location, the beamforming gain will reduce from $M$ to $|\sum_{m=1}^{M} e^{j 2\pi\phi_m}|^2/M \geq |\sum_{m=1}^{M} \cos(2\pi \phi_m)|^2/M \geq M \cos^2(2\pi\mu) \geq M/2$. Hence, in the worst case, the beamforming gain is reduced by 3\,dB, independently of $M$. This might seem counterintuitive, the beamwidth becomes narrower as $M$ increases, but is explained by the fact that we need be further away before the emitted signal takes the form of a beam. In conclusion, we never need estimate the channel more often than it takes to move $1/8$th of a wavelength and usually much less frequently (the chain of inequalities is very conservative).

\subsection*{mmWave: blessing and curse of attenuation and directivity}

The measuring and modeling of mmWave channels have received considerable attention, leading to a solid understanding of how these channels differ from sub-6\,GHz channels \cite{Rappaport2015} and extensions of the 3GPP channel models to support carrier frequencies from 0.5 to 100\,GHz (see 3GPP TR 38.901). We first consider the large-scale fading. Recalling the Friis transmission equation, the smaller wavelength $\lambda$ directly increases the path-loss proportionally to $\lambda^{-2}$. This is due to the implicit assumption of fixed-gain antennas whose effective area is proportional to $\lambda^{2}$. Hence, it can be overcome by using fixed-area antennas, which become increasingly directional with a gain proportional to $\lambda^{-2}$, or using an array of fixed-gain antennas whose total effective area is the same as on lower frequencies. The latter gives the flexibility to change the directivity of the array by beamforming, which is highly desirable in mobile communications. The feasibility of communicating at a high rate in LoS, benefiting from the wide available bandwidth, also over long distances has been exploited using high-gain directional antennas.

The total beamforming gain of a communication link is the product of the beamforming gains at the transmitter and receiver. Instead of deploying a huge array at one side of the link, the same total beamforming gain can be achieved by deploying substantially smaller arrays at both sides. For example, instead of having 1000 BS antennas to serve single-antenna UEs, we can have 100 BS antennas and 10 antennas per UE. This also opens the door to explore systems with massive arrays at both sides \cite{doubly_massive}.

The Fresnel zone defines the region around the LoS path that should be non-obstructed to avoid severe signal losses. Its radius, at a point located at distances $d_1$ and $d_2$ from the two ends of the link, respectively, is given by $\sqrt{\lambda\cdot d_{1}\cdot d_{2}/(d_{1}+d_{2})}$. At 38\,GHz, the Fresnel zone has $\approx0.5$\,m radius for a communication distance of 100\,m. For shorter distances and higher frequencies, it goes down to the cm-range. Hence, the Fresnel zone can be obstructed by small objects, leading to abrupt channel variations even when the transmitter and receiver are fixed. In mobile access, the signal strength will fluctuate rapidly as the obstruction changes. In combination with highly directive antennas, this calls for antenna arrays deliberately capturing reflections, or fast electronic beam-switching to reflected paths, if they are available.

At mmWave frequencies, many objects behave as full blockers, including humans \cite{Gustafson2012}, and there is less diffraction. Specific frequencies suffer from absorption by gases with colliding resonance frequencies, such as 60\,GHz for oxygen. Losses $>40$\,dB have been measured through a window, which is substantially higher than for sub-6\,GHz waves. The outdoor-to-indoor coverage is therefore rather limited in mmWave bands. Outdoors, significant losses through foliage have been also observed \cite{Rappaport2015}. Rain will cause higher attenuation with increasing frequencies, but the impact on the link budget is rather small. A consequence of these unfavorable propagation effects is that the link budget is worse in mmWave bands than at sub-6\,GHz, even if we let the physical size of the BS antenna array be the same in both bands.

The small-scale fading will also be considerably different with only one-bounce reflected paths actually contributing. 
The reflected paths may allow communication in case the LoS is blocked. Since the small-scale fading changes substantially when moving 1/8th wavelength, 10 times faster channel variations occur at 30\,GHz than at 3\,GHz when moving at the same speed, which calls for 10 times more frequent channel estimation. This might be less of an issue in practice; the coverage area of mmWave BSs is rather small, thus only low-mobility UEs will likely connect to them.

\section*{Difference II: Hardware implementation}

In mMIMO, an evident concern is the implementation complexity of  the digital baseband and analog/RF hardware. Technology scaling has fueled an impressive progress in wireless communication systems and is essential to process many antenna signals. 

The flexibility offered by full digital beamforming leads to the highest theoretically achievable performance, while hybrid analog-digital beamforming schemes are explored to enable hardware reuse over antenna paths, by having a mixed RF signal chain from the antennas to the digital baseband. 
However, neither the digital processing nor the data converters are a complexity hurdle, although those are the stages where hybrid beamforming primarily induces simplifications. It is the high-speed interconnect that is a bottleneck in the realization of integrated systems.

Next, we concisely discuss the key hardware sub-systems in mMIMO processing: the digital baseband and data converters, the RF and analog sub-systems, and the interconnection of the many antenna signals.

\subsection*{DSP and data converters: lean processing suits the system}

We distinguish three main parts in mMIMO DSP:
\begin{enumerate}
\item	
The outer modem applying error-correction coding on each data stream individually. Its complexity is not impacted by mMIMO.
\item
Central processing performing decoding and transmit beamforming. Operations on large matrices can be implemented efficiently in hardware when exploiting the nearly orthogonal user channels \cite{Prabhu2017}.
\item
Antenna signal processing by which we mean the DSP performed on signals connecting data converters to the central processing. Their overall complexity scales with the number of full (RF and analog) front-end chains from antennas to digital baseband. This part may be dominant in terms of operations/second, but can be implemented at low resolution \cite[Sec.~6]{massivemimobook}. 

\end{enumerate}
Taking advantage of scaled CMOS technology and system-level opportunities, efficient DSP implementations are feasible in both sub-6\,GHz and mmWave bands.

Data converters are a potential bottleneck in hardware complexity in multi-antenna processing. Low-power architectures have rendered this objection obsolete for BSs. Analog-to-digital converter (ADC) cores achieve figures-of-merit in terms of energy consumption per conversion step (cs)
in the order of $30\,\textrm{fJ}/(f_{s}2^\textrm{ENOB})$, where ENOB stands for ``effective number of bits.'' For each bit reduction in resolution, the ADC power is basically halved. For low resolutions even $10$\,fJ/cs has been reported \cite{Vanderplas2008}. 

ADCs in 4G systems (sub-6\,GHz) require a resolution $\textrm{ENOB}>10$, owing to the dynamic range requirement imposed by the combination of OFDM, MIMO, and high-order constellations. mMIMO in realistic conditions is expected to work well with $\textrm{ENOB}=5$ \cite[Sec.~6.4.1]{massivemimobook}. Hence, the ADC power consumption in a 128-antenna BS can be lower than in a conventional 8-antenna system. The actual power consumption of an integrated ADC may be a factor of 2-4 higher than in theory \cite{ExpertOpinions} to account for voltage regulators, input buffering, and calibration. Still, an individual converter may consume less than 1\,mW. A few hundred of them is hence negligible in the total power budget of a BS.

In mmWave systems, a multi-GHz bandwidth requires ADCs which, for physical reasons, consume an order of magnitude more power than their counterparts for sub-6\,GHz systems, considering similar linearity requirements. When equipping UEs with antenna arrays, the ADC power consumption will impact hybrid beamforming trade-offs.

\subsection*{Generating, phase shifting, and amplifying RF signals: divide and conquer?}

The synthesis of RF frequencies is challenged by strict constraints on phase-noise and error vector magnitude (EVM). These requirements are tougher to meet at mmWave compared to sub-6\,GHz. The oscillator efficiency is highly influenced by the ratio of the operating frequency over the channel spacing and the Q-factor of the resonator. For the former, one may assume channel spacing to go up with operating frequency. The Q-factor of the resonator typically drops at mmWave, resulting in a lower oscillator efficiency \cite{ExpertOpinions}.

Hybrid beamforming implemented with analog phase-shifting on the RF signals maximizes hardware reuse. For mmWave systems, the phase-shifters need to be able to settle fast to sustain communication when the LoS path is disrupted.
The realization of precise phase-shifting is difficult at high frequencies and may incur considerable power overhead. Hence, implementing the phase-shifting in the analog baseband may be preferred.

The power amplifier (PA) constitutes the most power-hungry component in RF transceivers. Linear PAs are required for the 5G broadband transmission schemes. mMIMO systems benefit from reduced output power requirements, both for the entire array and per antenna \cite[Sec.~5.2]{massivemimobook}. Their combined complexity (cost) and power will decrease with a growing number of antennas, with diminishing returns.

A sub-6\,GHz PA operating at 6\,dB back-off can achieve a power added efficiency (PAE) of 18\% \cite{Reynaert2016}. mmWave PAs need to rely on power combining, which introduces extra losses. Moreover, the lower gain at these frequencies calls for higher DC drive currents \cite{ExpertOpinions}. CMOS PAs achieve PAE<\,10\% at 6\,dB back-off. 

Operating PAs closer to saturation has been suggested for mMIMO systems in order to increase power-efficiency. However, this may infringe on the specification for EVM and out-of-band radiation as coherent combining of the non-linear distortion will occur in scenarios with a few dominant beams \cite{Larsson2018PA}, as expected in mmWave.

\subsection*{Interconnect is the main implementation challenge}

mMIMO systems process a large number of antenna signals.
Connecting these signals constitutes the main hardware implementation challenge. For sub-6\,GHz systems, in order to bring all individual signals to the DSP level, a balanced approach with partly distributed processing can circumvent the bottleneck \cite{Prabhu2017}.

\begin{figure}
        \centering
               \includegraphics[width=.4\columnwidth]{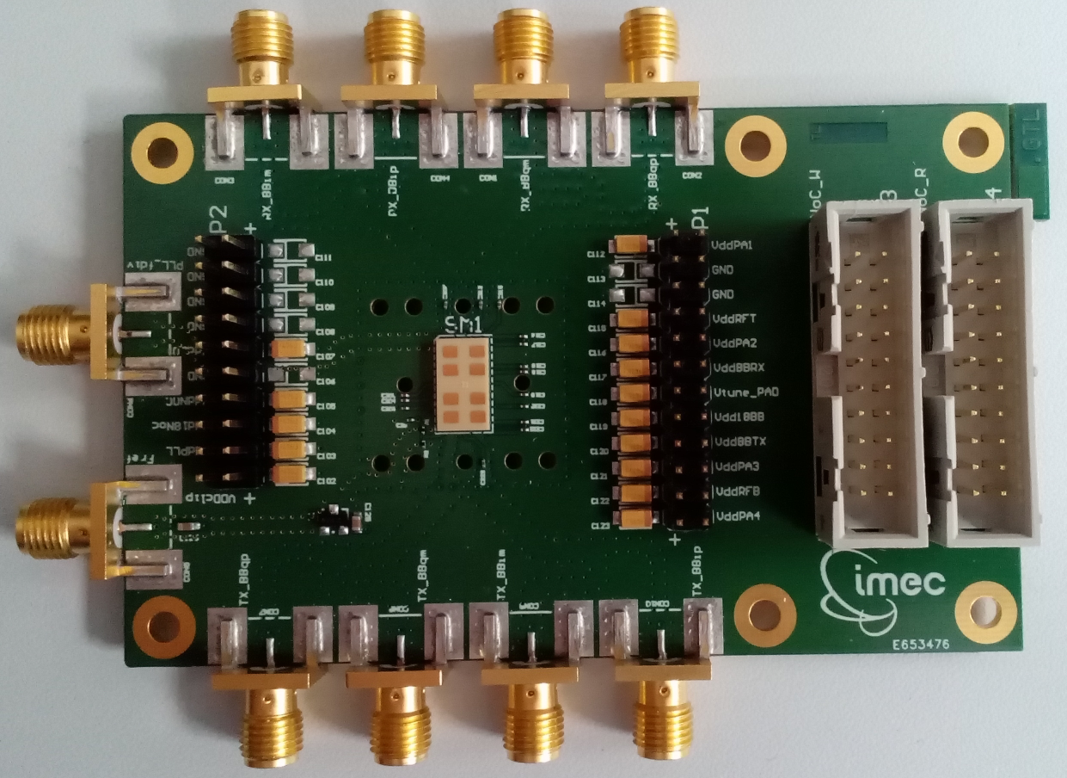} 
        \caption{mmWave module hosting a 4-antenna transceiver IC co-integrated with patch antennas (two patches for each of the four antenna paths). The size is 5.4 x 9.2\,mm. Courtesy of imec.}
        \label{figure:mmWavemodule}  
\end{figure}

At mmWave, the connections to the antennas become extremely lossy since micro-strip lines behave as antennas, giving losses of several\,dB/cm at 60\,GHz for different integration materials \cite{Brebels2014}. Matching of components is challenging \cite{ExpertOpinions}. Systems will only benefit from more antennas if these can be integrated in a very compact way, urging a co-design of chips, antennas, and package, as illustrated in Figure~\ref{figure:mmWavemodule} for the transceiver described in \cite{Mangraviti2016}. 

Unfortunately, hybrid beamforming does not relax the requirement of connecting mmWave signals to antennas. Oscillator distribution at mmWave frequencies also faces severe challenges; interconnects are the main bottleneck to exploit the high bandwidth through the integration of many small antennas.

\section*{Difference III: Signal processing algorithms}

The major differences in channel propagation and hardware implementation have fundamental implications on the algorithms needed for channel estimation, beamforming, and resource allocation. 

\subsection*{Opportunities for efficient channel estimation}

The number of channel coefficients grows linearly with the number of antennas at the BS and UE. To have an approximate idea of the computational burden, consider a system with 200 BS antennas and 20 spatially multiplexed single-antenna UEs. Consider OFDM with $1024$ subcarriers and channels that are constant over $12$ subcarriers. There are $3.4 \cdot 10^5$ complex scalar coefficients, which amounts to $6.8 \cdot 10^6$ estimates/second if a channel coherence time of 50\,ms is assumed. These numbers increase if there are more antennas, more subcarriers, and/or shorter coherence time.

At sub-6\,GHz, there is generally multi-path propagation caused by a multitude of scattering clusters. The channel coefficients are correlated across antennas, but this can only be utilized to marginally improve the estimation quality, at the cost of substantially higher complexity \cite[Sec.~3]{massivemimobook}. Nevertheless, the estimation can be conveniently implemented/parallelized in hardware \cite{Prabhu2017} and the estimation overhead is small when operating in time-division-duplex (TDD) mode and exploiting channel reciprocity to only send uplink pilots \cite{MLYN2016book}.

At mmWave, the channel can potentially be parameterized (considering a phase-synchronized array with a known angular array response) because it consists of a (potential) LoS path and few one-bounce reflections. Instead of estimating the individual channel coefficients, a few angular channel coefficients can be estimated to acquire the entire channel, leading to greatly reduced complexity. When a single data-stream is to be sent, it suffices to estimate the dominant angle-of-arrival/departure, but also reflections can be taken into account. However, if hybrid beamforming is used, the phase-shifters create a very directional ``vision'' and only channel components in that fall into the analog beams can be estimated. To discover new UEs, track channel variations, or keep the connection when the LoS path is blocked, beam-sweeping is needed (i.e., the channel must be estimated in many different directions to identify the preferable ones). This procedure increases the overhead from CSI acquisition, which grows with the number of antennas since the beams become narrower.

While TDD operation is preferable at sub-6\,GHz mMIMO, in mmWave bands frequency-division-duplex (FDD) may be equally good since the channel-describing angular parameters are reciprocal over a wide bandwidth.

\begin{figure}
        \centering
        \begin{subfigure}[b]{\columnwidth} \centering 
                \includegraphics[width=.7\columnwidth]{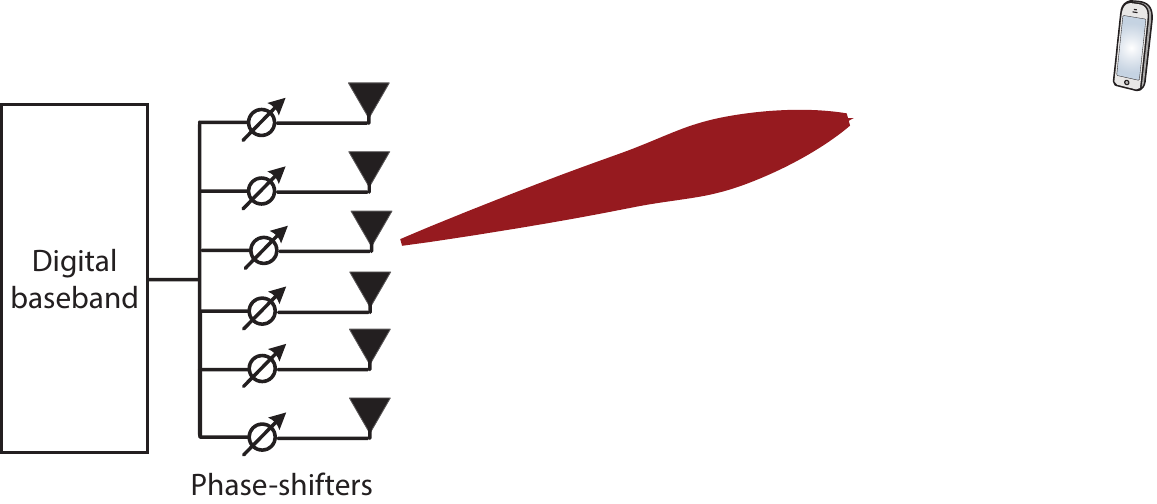} 
                \caption{Analog beamforming: Only one beam is created for the entire frequency band, which is sufficient for LoS beamforming.}  
                \label{figure:analog_beamforming}
        \end{subfigure}  \vskip5mm
\begin{subfigure}[b]{\columnwidth} \centering 
                \includegraphics[width=.7\columnwidth]{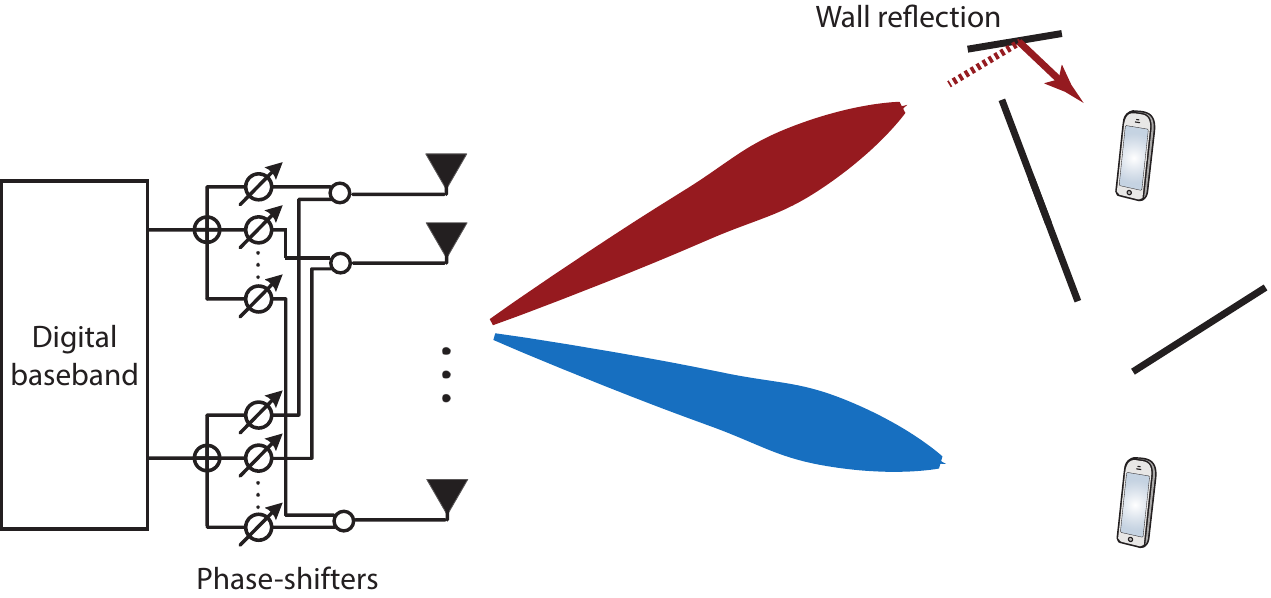} 
                \caption{Hybrid beamforming: A few beams are created by the analog phase-shifters and the digital baseband can create superpositions of these beams to adapt to multi-path and frequency-selective fading, which gives a limited  flexibility for handling NLoS scenarios.} 
                \label{figure:hybrid_beamforming}
        \end{subfigure}  \vskip5mm
        \begin{subfigure}[b]{\columnwidth} \centering
                \includegraphics[width=.75\columnwidth]{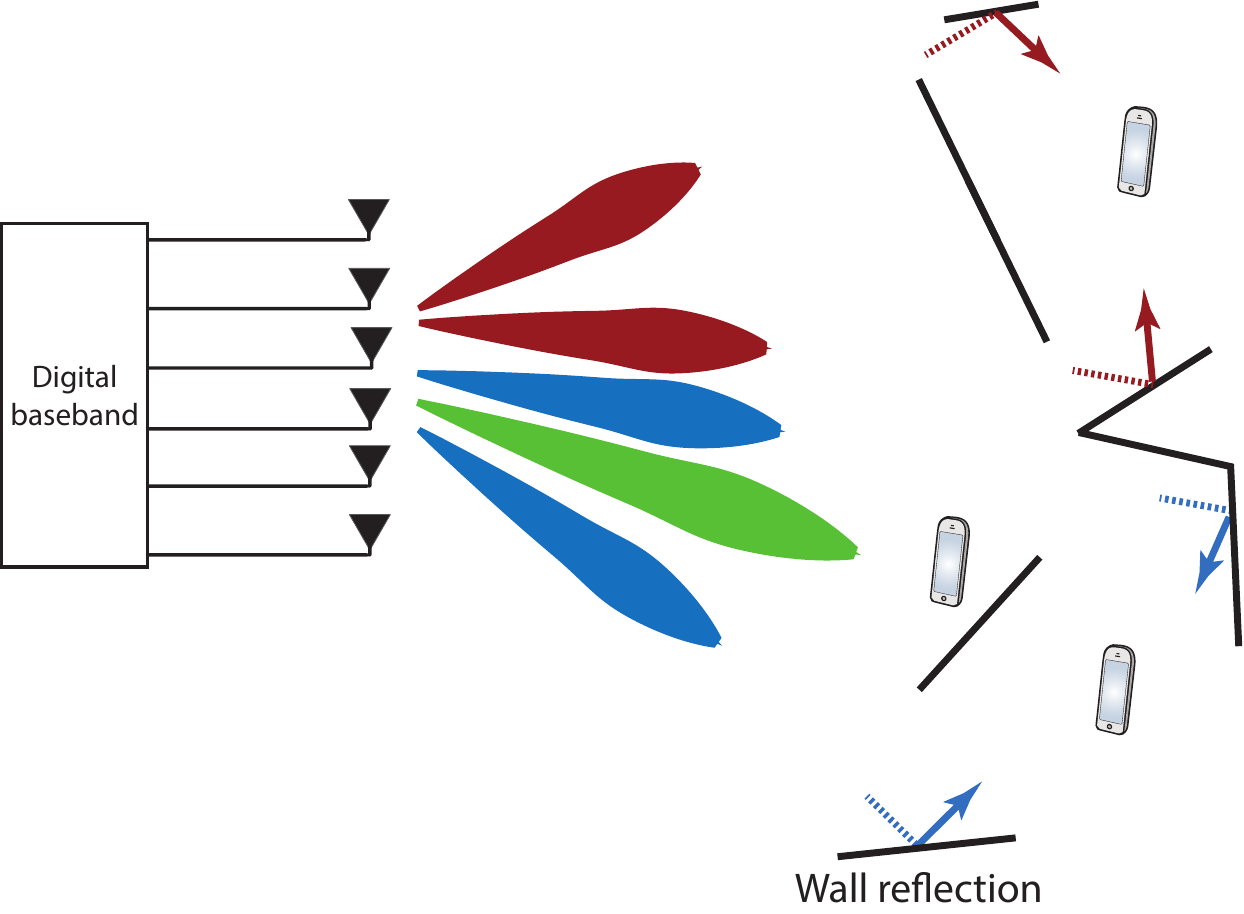} 
                \caption{Digital beamforming: Full flexibility to create a superposition of any number of beams and adapt the beams to multi-path and frequency-selective fading.} 
         \label{figure:digital_beamforming} 
        \end{subfigure} 
        \caption{The hardware implementation of beamforming determines the flexibility in handling difficult propagation scenarios, which in turn has implications on the use cases. The system is viewed from above and different colors represent different signals.}
        \label{figure:beamforming}  
\end{figure}

\subsection*{Choosing between analog, hybrid, and digital beamforming}

Current hardware can realize full digital beamforming at sub-6\,GHz, while hybrid analog/digital beamforming is a potential design-simplification at mmWave. With analog transmit beamforming, a phase-shifted version of the same signal is transmitted from all antennas. This leads to a signal beam directed in a particular angular direction; see Figure~\ref{figure:analog_beamforming}. If multiple UEs are multiplexed, one set of analog phase-shifters (connected to a separate input from the baseband) is needed per UE; see Figure~\ref{figure:hybrid_beamforming}. This is hybrid beamforming in a nutshell. The number of UEs cannot be larger than the number of baseband-inputs, but digital precoding can be used to assign a mix of the UEs' signal to each input.

In contrast, full digital beamforming can send any signal from any antenna. This flexibility can be exploited at sub-6\,GHz frequencies to deliver high beamforming gain in rich multi-path environments, as illustrated in Figure~\ref{figure:digital_beamforming}. The digital flexibility is evident in multi-user scenarios, where the antennas should transmit a superposition of many beams per UE, different beams per subcarriers (due to frequency-selective fading), and multiplex many UEs on the same time-frequency resource slot. Analog beamforming cannot adapt the beam to multi-path propagation and frequency-selective fading, while hybrid beamforming has only a limited ability to do that since the phase-shifters create a set of fixed beams that the digital precoding needs to be based on.

mmWave systems have lower user multiplexing capability if implemented with hybrid beamforming, since the number of UEs is limited by the number of sets of phase-shifters.
However, even analog beamforming (as in Figure~\ref{figure:analog_beamforming}) suffices for single-user communications over wide bandwidths. To illustrate this fact, we consider a LoS channel with five reflections. The center frequency is 60\,GHz and we use the method in \cite[Sec.~7.3.2]{massivemimobook} to compute the array response for different frequencies. We consider  $32\times 32$, $64\times 64$, and $128\times 128$ planar arrays with half-wavelength-spaced antennas. 

\begin{figure}
        \centering
                \includegraphics[width=.8\columnwidth]{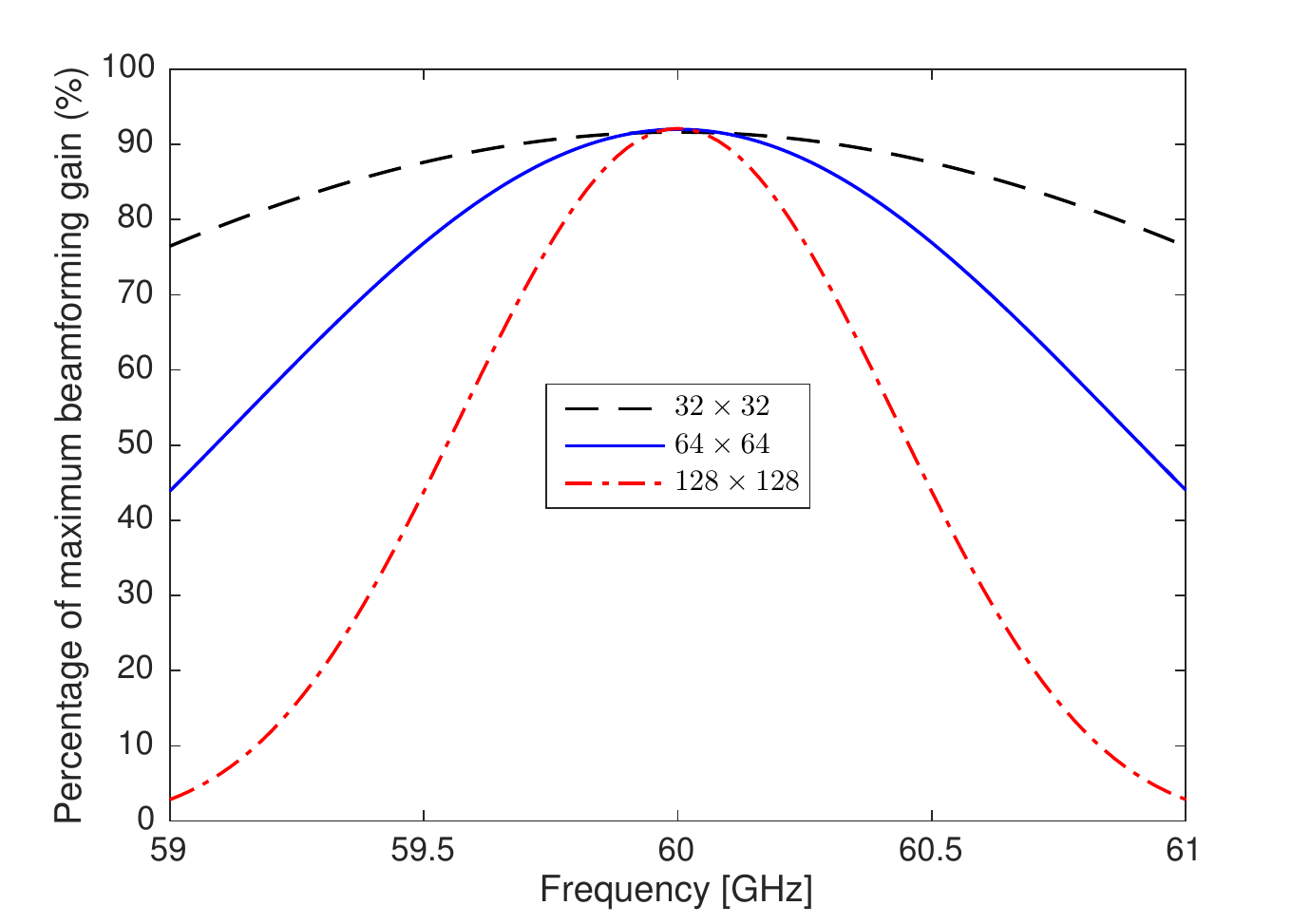} 
        \caption{The fraction of the maximum beamforming gain that is achieved at different frequencies when analog beamforming is used. The phases are selected to maximize the inner product with the array response at the center frequency.
        There is a LoS path with azimuth angle $\pi/4$ and elevation angle $-\pi/4$ to the array, measured from the boresight.
        There are also five reflections, with the azimuth angles $\pi/6$, $\pi/3$,$\pi/4$, $\pi/4$, $\pi/12$ and the corresponding elevation angles $-\pi/5$, $-\pi/5$, $-\pi/6$, $-\pi/12$, and $-\pi/6$. The total gain of the LoS equals the gain of all the reflections.}
        \label{figure:beamsquinting}  
\end{figure}

The maximum beamforming gain is equal to the number of antennas and is achievable by digital beamforming. Figure~\ref{figure:beamsquinting} shows the percentage of the maximum beamforming gain that is obtained by analog beamforming at different frequencies around the center frequency. It starts at 90\% since no analog beamforming matches the array response when there are reflections. Nevertheless, if the bandwidth is 400\,MHz, 80-90\% of the maximum beamforming gain can be achieved in the entire band by analog beamforming. If the bandwidth continues to grow, the beamforming gain drops since the beamforming is optimized for the center frequency. This is known as the beam-squinting effect. The gain loss is particularly severe for larger arrays, since the beams are narrower. With the $32\times 32$ array, more than 75\% of the maximum gain can be achieved over a 2\,GHz bandwidth, while the  gain drops quickly for the $128\times 128$ array.

\subsection*{Network deployment for good coverage}

Several differences between sub-6\,GHz and mmWave arise in network planning and resource allocation. At sub-6\,GHz, BSs ensure both outdoor and indoor coverage, and support to high-mobility UEs. Although the BS positioning is important, it is not as crucial as the adoption of interference-management procedures, such as advanced digital beamforming techniques that deal with inter-cell interference and pilot contamination \cite[Sec.~4]{massivemimobook}. 

In contrast, at mmWave frequencies, given the blockage effects, very careful deployment planning is needed to provide coverage to an intended area. Interference is less important, but ensuring wide-area coverage, without coverage holes, may require a large number of BSs.

\section*{Use-cases: Different solutions for different cases}

Although the data traffic increases by 30-40\% annually, contemporary macro-cells only need to serve one or a few UEs at any given time instant. The reason is that the networks have been gradually densified. In traffic-intense areas, the inter-BS distance is in the order of 100\,m, rendering further densification questionable from a practical and cost perspective. Hence, the number of simultaneous UEs is likely to grow rapidly in the future. The new use cases of \emph{Ultra reliable low latency communication} (URLLC) and \emph{Massive machine-type communications} (mMTC) to support diverse Internet-of-things (IoT) applications are two drivers towards this change. A world with sensors everywhere, autonomous cars, drones and social robots, and augmented-reality applications will require a network infrastructure that supports 100 times higher capacity than today.

The key use-cases and the propagation scenarios are summarized in Table~\ref{table:use-cases}. One cannot separate these aspects since a technology can be a perfect fit for a use-case in one scenario, but infeasible in another scenario; for example, mMIMO in mmWave bands can provide unprecedented data rates in LoS scenarios, but is less suited for outdoor-to-indoor communications. To deliver all the necessary services, we need to evolve the networks in two respects: 1) improve the macro-cell BSs to handle many simultaneous UEs; 2) deploy short-range BSs that offload traffic in hotspots.

\begin{table} \vspace{-5mm}
\begin{center}
        (a) Feasibility and suitability of mMIMO in different uses cases.

    \begin{tabular}{ | p{4.4cm} | p{4.7cm} |  p{4.7cm} |}
    \hline
    \textbf{Use case} & \textbf{mMIMO in sub-6\,GHz} & \textbf{mMIMO in mmWave} \\ \hline \hline
    
    Broadband access & High data rates in most propagation scenarios (e.g., $\sim$100\,Mbit/s/user using $40$\,MHz of bandwidth), with uniformly good quality-of-service  & Huge data rates (e.g., $\sim$10\,Gbit/s/user using several GHz of bandwidth) in some propagation scenarios (see below). \\ \hline
    
    IoT, mMTC & Beamforming gain gives power-savings and better coverage than legacy networks & Not fit for low data rate applications, which will incur significant power overhead  \\ \hline
    
    URLLC & Channel hardening improves reliability over legacy networks & Difficult since propagation is unreliable due to blockage \\ \hline
    
    Mobility support & Same great support as in legacy networks  & Very challenging, but theoretically possible \\ \hline
    
    High throughput fixed link & Narrow beamforming is possible with 100 antennas, 20\,dB beamforming gain is achievable; only array size limits the gain & Beamforming gain is possibly higher than at sub-6\,GHz, since more antennas fit into a given area, but the gain per antenna is smaller \\ \hline
    
    High user density & Spatial multiplexing of tens of UEs is feasible and has been demonstrated in field-trials & Same capability as at sub-6\,GHz in theory, but practically limited if hybrid implementation is used \\ \hline
        \end{tabular}

        \vspace{.7cm}
        (b) Feasibility and suitability of mMIMO in different propagation scenarios. 
    \begin{tabular}{ | p{4.4cm} | p{4.7cm} |  p{4.7cm} |}
    
    \hline
    \textbf{Propagation scenario} & \textbf{mMIMO in sub-6\,GHz} & \textbf{mMIMO in mmWave} \\ \hline \hline
    
    Outdoor-to-outdoor, indoor-to-indoor communication & High data rates and reliability (see above) in both LoS and NLoS scenarios & Huge data rates (see above) in LoS hotspots, but unreliable due to blockage phenomena \\ \hline
    
    Outdoor-to-indoor communication & High data rates and reliability (see above) & Limited due to higher propagation losses \\ \hline
    
    Backhaul/fronthaul links & Can multiplex many links, even in NLoS, but relatively modest data rates per link & Great for LoS links, particularly for fixed antenna deployments, but less suitable for NLoS links \\ \hline
    
    Operational regime & Mainly interference-limited in cellular networks, due to high SNR from beamforming gains and substantial inter-user interference & Mainly noise-limited in indoor scenarios, due to huge bandwidth and limited inter-cell interference, but can be interference-limited outdoors \\ \hline
    
    \end{tabular}
\end{center} \vspace{-4mm}
\caption{Feasibility and suitability of mMIMO at sub-6\,GHz and mmWave for different use cases and propagation scenarios.} \label{table:use-cases}
\end{table}

\begin{itemize}
\item[-] 
\textbf{Macro-cells:} mMIMO at sub-6\,GHz is ideal for delivering higher throughput in macro-cells than in legacy networks. As noted in Table~\ref{table:use-cases}, the cell-edge and outdoor-to-indoor coverage are improved by the beamforming gain: the received useful signal power grows proportionally to the number of antennas, whereas the (average) interference power at other locations remains the same due to non-coherent combination. While network densification does not improve average cell-edge conditions, since both the desired and interfering signals become larger, the beamforming gain does improve for the cell-edge UEs by only increasing the desired signals. Hence, it can be utilized to provide uniformly high quality-of-service throughout the cell. With 40\,MHz bandwidth and 3\,bit/s/Hz, data rates of 120\,Mbit/s can be achieved per UE. By multiplexing 20 UEs, the cell throughput becomes 2.4\,Gbit/s.

mMIMO at sub-6\,GHz offers the same support for user mobility as other technologies operating in that band, and high-mobility support has been demonstrated in field-trials \cite{Harris2017a}.
 
Since the purpose of using mmWave bands is to have 10-100 times more bandwidth than at sub-6\,GHz, the link budget will be reduced by 10-20\,dB (assuming the same output power and effective antenna area at the BS). When combined with the fact that outdoor-to-indoor propagation is rather limited and the signals are easily blocked, a huge number of BSs, relays, and/or reflective surfaces would be needed to guarantee wide-area coverage. The mmWave band is, however, attractive for providing fixed wireless access over large areas, since the BSs and UEs can then be deployed to guarantee LoS-like conditions.

\item[-]
\textbf{Hotspots:} Auditoriums, caf\'es, airports, and stadiums are examples of hotspots, where the data traffic is very high in a physically small area. To offload the macro-cells, WiFi is most frequently used in these places, but WiFi neither supports mobility nor high user loads. These issues can be resolved by using mMIMO at sub-6\,GHz (an array need not be larger than a television screen), but since LoS propagation dominates in hotspots, mmWave mMIMO is a more suitable solution. In hotspots, a decent signal-to-noise ratio (SNR) can be achieved over a huge bandwidth, thanks to the short propagation distances, leading to extreme throughput. For example, with 1\,GHz of bandwidth, a spectral efficiency of 1\,bit/s/Hz is sufficient to achieve 1\,Gbit/s per data stream. With more spectrum and/or higher spectral efficiency, 10\,Gbit/s is within reach. This is a key use-case for mmWave technology. Spatial multiplexing of UEs can be implemented using hybrid beamforming, as illustrated in Figure~\ref{figure:hybrid_beamforming}, if the UEs are in LoS. Since the channels evolve twenty times faster when going from, say, 3\,GHz to 60\,GHz carrier frequency, mmWave hotspots can easily support pedestrian movement, while higher speeds are more challenging.

\end{itemize}

\subsection*{What if extra hardware came at no cost?}

Suppose the hardware and signal processing come for free and work perfectly, how large an array could eventually be useful?

In an environment without significant mobility, very large numbers of users may be spatially multiplexed. In \cite[Sec.~6.1]{MLYN2016book}, one sub-6\,GHz case study establishes the feasibility of providing (fixed) wireless broadband service to 3,000 homes, using a BS with 3,200 antennas (which at 2\,GHz requires an array of around $4\times4$\,m). By jointly increasing the number of antennas and UEs, the total radiated power per BS and rate per UE can be kept constant. 

The number of UEs that can be spatially multiplexed per BS is determined by the number of samples per channel coherence time-frequency block and the number of BS antennas. An outdoor network that supports high mobility has a few hundred samples per coherence block when operating at sub-6\,GHz, giving room to orthogonal resources for channel estimation to a few hundred UEs. This number is inversely proportional to the carrier frequency \cite[Sec.~2]{MLYN2016book}, leading to an order-of-magnitude of fewer samples in mmWave bands.

\begin{figure}
        \centering
        \begin{subfigure}[b]{\columnwidth} \centering 
                \includegraphics[width=.8\columnwidth]{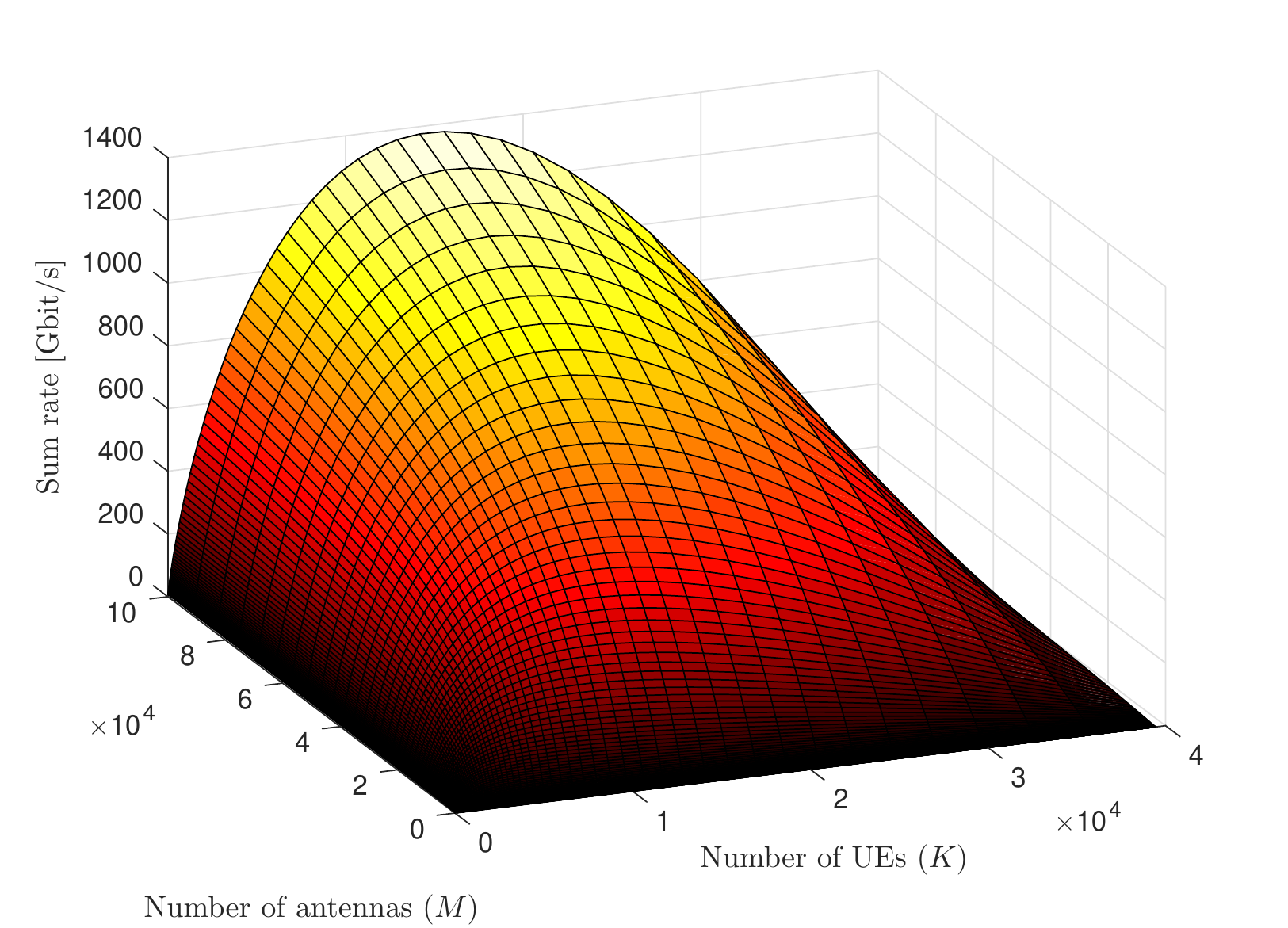} 
                \caption{3\,GHz carrier frequency  and 50\,MHz bandwidth.}  
                \label{figure:extrememultiplexing_A}
        \end{subfigure}  \vskip5mm
        \begin{subfigure}[b]{\columnwidth} \centering
                \includegraphics[width=.8\columnwidth]{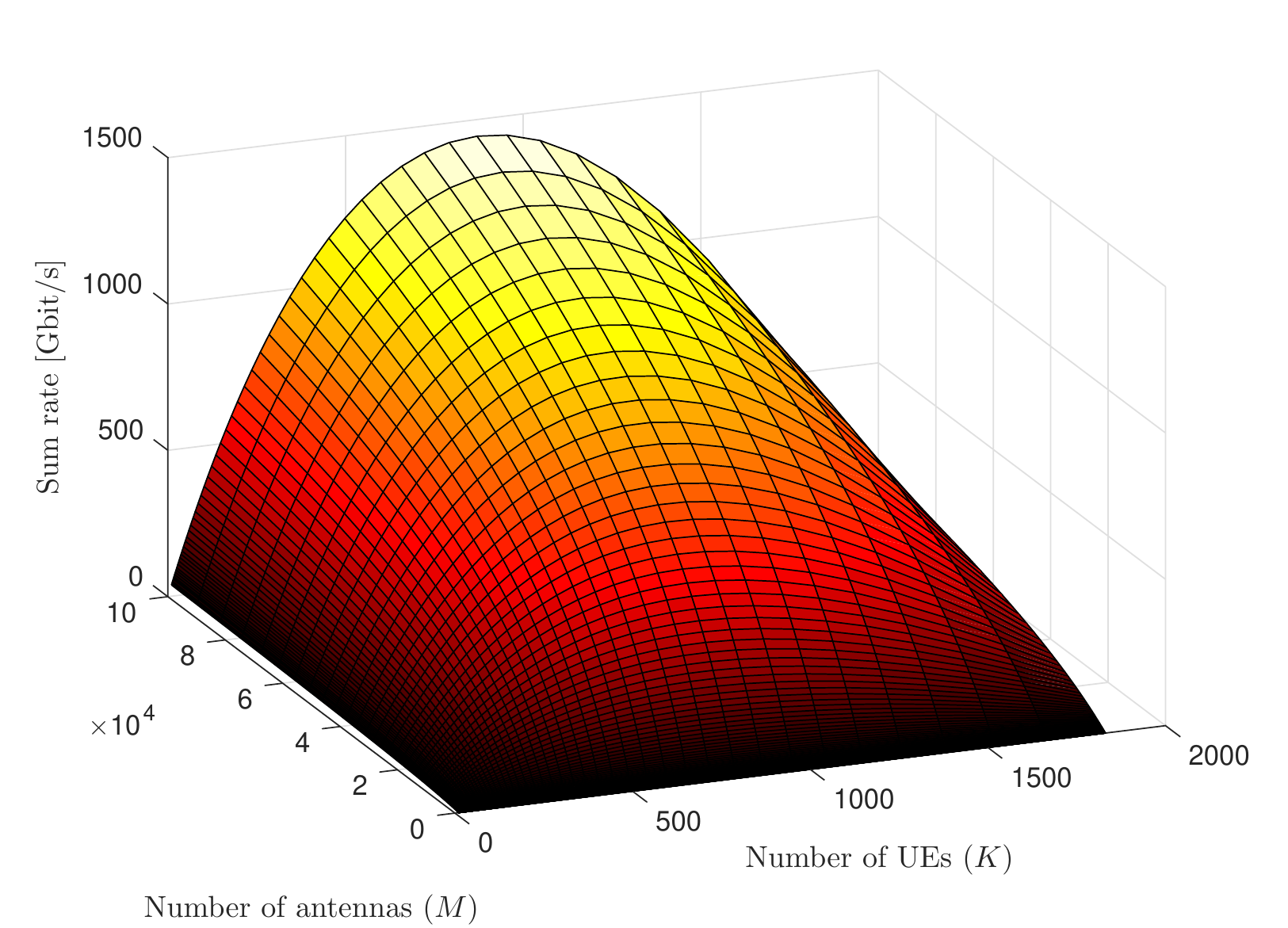} 
                \caption{60\,GHz carrier frequency and 1\,GHz bandwidth.} 
                \label{figure:extrememultiplexing_b} 
        \end{subfigure} 
        \caption{Downlink sum rate that is achieved when operating at different carrier frequencies using TDD and digital beamforming, as a function of the number of BS antennas. The uplink SNR to each receive-antenna is 20\,dB when operating at 3\,GHz with 50\,MHz bandwidth and scaled accordingly when operating at 60\,GHz with 1\,GHz bandwidth to keep the transmit power fixed. The downlink transmission uses 100 times more power than the uplink pilot transmission. Closed-form rate expressions from \cite[Sec.~3.3]{MLYN2016book} were used to generate the figures. To make the signal processing complexity scalable, maximum ratio transmission and channel estimation based on uplink pilots are assumed.}
        \label{figure:extrememultiplexing}  
\end{figure}

In an environment without significant mobility, very large numbers of UEs may be multiplexed \cite[Sec.~6.1]{MLYN2016book}. Consider, for example, a festival taking place in Central Park, Manhattan. This large park is surrounded by skyscrapers, where BS antennas can be mounted to provide LoS conditions. Eventually, only measurements can determine the channel coherence, but assume for the sake of argument a coherence block of 100\,ms by 400\,kHz when operating at 3\,GHz. The coherence time reduces to 5\,ms when operating at 60\,GHz.

Figure~\ref{figure:extrememultiplexing} shows the downlink sum rate when operating at these frequencies, as a function of the number of antennas and UEs, and assuming that fully digital reciprocity-based beamforming is used in all cases.
The sum rate grows monotonically with the number of antennas, as expected. The highest values on the curves are 1.38\,Tbit/s at 3\,GHz (with 50\,MHz bandwidth) and 1.44\,Tbit/s at 60\,GHz (with 1\,GHz bandwidth), which are nearly the same. The huge difference is that the peak values are achieved by multiplexing 14,000 or 870 UEs, respectively. This corresponds to allocating 35\% and 44\% of the coherence blocks to uplink pilots, respectively. The mmWave setup delivers 1.66\,Gbit/s per UE, while the sub-6\,GHz setup only delivers 99\,Mbit/s per UE, but compensates by serving extremely many UEs. This exposes the fundamental operational difference; the huge bandwidth in mmWave bands allows for high per-UE rates, while the longer coherence time at sub-6\,GHz allows for spatial multiplexing of more UEs. Which solution that is preferable depends on the data traffic characteristics of the future, but why not deploy both?

The maximum number of antennas was 100,000 in this futuristic simulation. Assuming 3\,GHz and half-wavelength-antenna-spacing, these antennas can be deployed in array of 31\,m$\times$31\,m. At 60\,GHz, this shrinks to 1.58\,m$\times$1.58\,m. Both setups can easily be deployed at the face of a skyscraper, so the size is not an issue. However, adequate implementation strategies are needed to cope with bottlenecks in connecting and processing the many signals.

\section*{Conclusions and the way ahead}

This paper has reviewed the major differences in mMIMO design for sub-6\,GHz and mmWave frequencies, concerning the propagation mechanisms, transceiver hardware, and signal processing algorithms. The impact on the various envisioned 5G use-cases has been explained, showing that both bands offer attractive propositions. 
Computational complexity is no longer a main bottleneck, but less considered factors, such as the interconnect of signals, both for central baseband processing and at mmWave to antennas, constitute potential bottlenecks. The technology is at a more advanced stage at sub-6\,GHz, yet challenges exist in both bands. Several intriguing questions remain unanswered: Will mmWave mMIMO be implemented with full digital beamforming? Which mMIMO features will be actually used in 5G networks? Will the multiplexing capabilities ever be pushed as high as illustrated in the Central Park example? How will data-traffic patterns and applications evolve? Whatever the answers will be, mMIMO will certainly play a paramount role in the shaping of future wireless networks in both bands.

While this article has substantiated how mMIMO offer unprecedented performance to end users, other applications are envisioned, such as the implementation of cloud-RAN through in-band wireless fronthauling \cite[Sec.~7.6]{massivemimobook}. The enormous amount of baseband data available in mMIMO systems can be also used to sense the environment; for example, estimate the amount of traffic on a road, count the number of persons in a room, or guard against intrusion in protected spaces. 
In conclusions, as far as mMIMO is concerned, the best is yet to come.


\begin{thebibliography}{99}
\bibitem{Marzetta}
T. L. Marzetta,
``Noncooperative cellular wireless with unlimited numbers of base station antennas,"
{\em IEEE Transactions on Wireless Communications,} vol. 9, no. 11, pp. 3590-3600, Nov. 2010.

\bibitem{Gao2015}
X. Gao, O. Edfors, F. Rusek, F. Tufvesson, ``Massive MIMO performance evaluation based on measured propagation data,"
{\em IEEE Transactions on Wireless Communications}, vol. 14, no. 7, pp. 3899-3911, July 2015.

\bibitem{Harris2017a}
P. Harris, S. Malkowsky, J. Vieira, E. Bengtsson, F. Tufvesson, W. B. Hasan, L. Liu, M. Beach, S. Armour, O. Edfors, "Performance Characterization of a Real-Time Massive MIMO System with LOS Mobile Channels," {\em IEEE Journal on Selected Areas in Communications}, vol. 35, no. 6, pp. 1244-1253, June 2017.

\bibitem{MLYN2016book}
T.~L. Marzetta, E.~G. Larsson, H.~Yang, and H.~Q. Ngo, \emph{Fundamentals of
  Massive MIMO}.  Cambridge University
  Press, 2016.

\bibitem{massivemimobook}
E. Bj\"{o}rnson, J. Hoydis, L. Sanguinetti,
``Massive MIMO Networks: Spectral, Energy, and Hardware Efficiency,"
{\em Foundations and Trends{\textregistered} in Signal Processing},
vol. 11, no. 3-4, pp. 154-655, 2017.

\bibitem{Rappaport2015}
T. S. Rappaport, G. R. MacCartney, M. K. Samimi and S. Sun, 
``Wideband Millimeter-Wave Propagation Measurements and Channel Models for Future Wireless Communication System Design," {\em IEEE Transactions on Communications}, vol. 63, no. 9, pp. 3029-3056, Sept. 2015.

\bibitem{doubly_massive}
S. Buzzi and C. D'Andrea, 
``Energy efficiency and asymptotic performance evaluation of beamforming structures in  doubly massive MIMO mmWave systems,"
{\em IEEE Trans. on Green Communications and Networking}, vol. 2, no. 2, pp. 385-396, June 2018.

\bibitem{Gustafson2012}
C. Gustafson and F. Tufvesson, ``Characterization of 60 GHz shadowing by human bodies and simple phantoms," {\em 6th European Conference on Antennas and Propagation (EUCAP)}, Prague, 2012, pp. 473-477.

\bibitem{Prabhu2017}
H. Prabhu, J. N. Rodrigues, L. Liu and O. Edfors, ``3.6 A 60pJ/b 300Mb/s 128×8 Massive MIMO precoder-detector in 28nm FD-SOI," {\em  IEEE International Solid-State Circuits Conference (ISSCC)}, vol.~60, pp.~60-61, San Francisco, CA, Feb. 2017.


\bibitem{ExpertOpinions}
Oral conversations with experts Dr. Bob Verbruggen, Xilinx, on ADCs, Prof.~Pietro Andreani, Lund University, on frequency synthesis, Prof.~Patrick Reynaert, KU Leuven, on power amplifiers, and Prof.~Dominique Schreurs, KU Leuven, on microwave circuits.


\bibitem{Reynaert2016}
P. Reynaert, Y. Cao, M. Vigilante and P. Indirayanti, ``Doherty techniques for 5G RF and mm-wave power amplifiers," {International Symposium on VLSI Design, Automation and Test (VLSI-DAT)}, pp. 1-2, Hsinchu, 2016.


\bibitem{Larsson2018PA} 
E. G. Larsson and L. Van der Perre, 
``Out-of-Band Radiation from Antenna Arrays Clarified", IEEE Wireless Communications Letters, to appear, 2018.

\bibitem{Vanderplas2008}
G. Van der Plas and B. Verbruggen, ``A 150 MS/s 133$~mu$W 7 bit ADC in 90 nm Digital CMOS," {\em IEEE Journal of Solid-State Circuits}, vol. 43, no. 12, pp. 2631-2640, Dec. 2008.

\bibitem{Brebels2014}
S. Brebels, A. A. Enayati, C. Soens, W. De Raedt, L. Van der Perre and G. A. E. Vandenbosch, ``Technologies for integrated mm-Wave antenna,.'' {\em The 8th European Conference on Antennas and Propagation (EuCAP 2014)} pp. 727-731, The Hague, Apr. 2014.

\bibitem{Mangraviti2016}
G. Mangraviti et al., ``A 4-antenna-path beamforming transceiver for 60GHz multi-Gb/s communication in 28nm CMOS,'' {\em 2016 IEEE International Solid-State Circuits Conference (ISSCC)},  pp. 246-247, San Francisco, CA, Feb. 2016.

\end{thebibliography}
\end{document}